\documentclass{article} 
\usepackage{amsmath,amssymb,amsbsy,amsfonts}
\makeatletter

          \@addtoreset{equation}{section}
       \makeatother
\allowdisplaybreaks 
\begin{document}
\title{\bf Lorentz covariant field theory on noncommutative spacetime
based on DFR algebra}%

\author{Yoshitaka {\sc Okumura},\\[3mm] 
{Department of Natural Science, 
Chubu University, Kasugai, {487-8501}, Japan}
}%
\date{}%
\maketitle
\begin{abstract}
Lorentz covariance is the fundamental principle of every relativistic field theory which insures consistent physical descriptions. Even if the space-time is noncommutative, field theories on it should keep Lorentz covariance.
In this letter, it is shown that
the field theory on noncommutative spacetime is  Lorentz covariant
if the noncommutativity emerges from the algebra of spacetime
operators described by Doplicher, Fredenhagen and  Roberts.
\end{abstract}



\thispagestyle{empty}
\section{Introduction\label{S1}}
In the past several years, field theories on the noncommutative(NC) spacetime
have been extensively studied from many different aspects. The motivation 
comes from the string theory which makes obvious that end points of 
the open strings trapped on the D-brane in the presence of 
 two form B-field background turn out to be noncommutative 
\cite{SJ}
and 
then the noncommutative supersymmetric gauge theories appear as the low energy effective theory of such D-brane \cite{DH}, \cite{SW}.
The noncommutativity of spacetime in recent surge 
is characterized by the algebra
$[x^\mu,\,x^\nu]=i\theta^{\mu\nu}$ where $\theta^{\mu\nu}$ 
is a real, anti-symmetric constant with dimension 2, which is reflected as the Moyal star product in field theories. 
According to this prescription, one can build NC version of 
scalar, Dirac and gauge theories. Thus, apart from the string theory,
the studies of NC field theories have been proposing
very interesting as well as reversely serious outcomes.
The NC scalar field theories are investigated in \cite{Filk}, 
\cite{SST},\cite{Micu}, \cite{GM}.
It was showed that the NC scalar theory with $\mit\Phi^4$
interaction is renormalizable and the parameter $\theta^{\mu\nu}$ 
doesn't receive the quantum corrections up to two loop order. 
However, since the Moyal star product contains an infinite series of 
$x^\mu$ derivatives, NC field theories are nonlocal.
The nonlocality especially in timelike noncommutativity $\theta^{0i}\ne0$ 
leads to the unitarity violation \cite{GM} and the difficulty of 
renormalizability. 
All these investigations are 
carried out under the condition of constant $\theta^{\mu\nu}$, 
which means that Lorentz covariance of theory is violated. \par
Doplicher, Fredenhagen and Roberts (DFR) proposed \cite{DFR} a new algebra of 
NC spacetime through consideration of the spacetime uncertainty
relations derived from quantum mechanics and general relativity.
In their algebra, $\theta^{\mu\nu}$ is promoted to an anti-symmetric tensor operator, which leads to the Lorentz covariant NC spacetime and
enables one to construct the Lorentz invariant NC field theories.
Carlson, Carone and Zobin (CCZ) \cite{CCZ} 
formulated the NC gauge theory by referring to the DFR algebra
in the Lorentz invariant way. In their formulation, fields in the theory
depend on spacetime $x^\mu$ and the NC parameter
$\theta^{\mu\nu}$. The action is obtained by integrating Lagrangian 
over spacetime $x^\mu$  as well as the NC parameter
$\theta^{\mu\nu}$.
The characteristic features of CCZ \cite{CCZ} are to set up the 6-dimensional
$\theta$ space in addition to $x$ space and define the action as the integration of Lagrangian over the $\theta$ space as well as $x$ space.
In this article, we will construct Lorentz covariant NC field theory by  taking the trace of ${\cal{L}}(\hat{x},\,\hat{\theta})$ over spacetime operator $\hat{x}^\mu$, and taking  the
matrix element of ${\cal{L}}(\hat{x},\,\hat{\theta})$ 
between $|\theta>$ and $<\theta|$,  rather than 
  the integration over $\theta$ space as in CCZ \cite{CCZ}. 
We can choose the spacelike noncommutativity ($\theta^{0i}=0$) after 
the appropriate Lorentz transformation and so any serious outcomes such as 
quantization and unitarity violation don't appear.
\section{Lorentz covariance of noncommutative field theory}
Lorentz covariance is the fundamental principle of every relativistic field theory which insures consistent physical descriptions such as causality, 
unitarity and so on. However, it hasn't been respected in the study of 
NC field theory so far. Doplicher, Fredenhagen and Roberts (DFR)
\cite{DFR}
first addressed this problem to propose a new algebra of NC
spacetime operator ${\hat x}_\mu$.
\begin{equation}
[{\hat x}^\mu,\,{\hat x}^\nu]=i\,{\hat \theta}^{\mu\nu},
\end{equation}
where ${\hat \theta}^{\mu\nu}$ is an antisymmetric tensor operator, not a constant considered so far. They further assumed 
\begin{equation}
[{\hat x}^\mu,\,{\hat \theta}^{\mu\nu}]=0,
\end{equation}
which leads to the commutativity between ${\hat \theta}^{\mu\nu}$
through the Jacobi identity
\begin{equation}
[{\hat \theta}^{\mu\nu},\,{\hat \theta}^{\sigma\rho}]=0.
\label{3.3}
\end{equation}
Equation \eqref{3.3} enables us to simultaneously diagonize the operator
${\hat \theta}^{\mu\nu}$.
\begin{equation}
{\hat \theta}^{\mu\nu}|\,\theta>={\theta}^{\mu\nu}|\,\theta>,
\label{3.4}
\end{equation}
where $|\,\theta>$ is a eigenstate and ${\theta}^{\mu\nu}$ is 
its specific eigenvalue.\par
Carlson, Carone and Zobin (CCZ) \cite{CCZ} 
formulated the NC gauge theory by referring to the DFR algebra
in the Lorentz invariant way. In their formulation, fields in the theory
depend on spacetime $x^\mu$ and the NC parameter
$\theta^{\mu\nu}$. The action is obtained by integrating Lagrangian 
over spacetime $x^\mu$  as well as the NC parameter
$\theta^{\mu\nu}$.
\begin{equation}
S=\int d^{\,4}x \,d^{\,6}\theta\; 
W(\theta){\cal L}(\phi(x,\theta),\partial^\mu 
\phi(x,\theta)),
\end{equation}
where Lorentz invariant function $W(\theta)$ is a weight function to render the $\theta$ integral finite. Following to CCZ, Kase, Morita, Okumura and Umezawa 
\cite{KMOU} reconsidered the Lorentz invariant NC field theory by  
pointing out the inconsistency of the c-number $\theta$-algebra
 and indicated that
the normalizability of the weight
function in Lorentz metric leads to the division of the $\theta$ space 
into two disjoint regions not connected by any Lorentz transformation
, so that the CCZ covariant moments formula holds in each region separately.
\par
The characteristic features of CCZ \cite{CCZ} are to set up the 6-dimensional
$\theta$ space in addition to $x$ space and define the action as the integration of Lagrangian over the $\theta$ space as well as $x$ space.
In this article, we take over the idea of the 6-dimensional $\theta$ space, but
don't  the integration over $\theta$ space. Let us pick up one specific point 
$\theta^{\mu\nu}$ in 
the 6-dimensional $\theta$ space that follows from Eq.\eqref{3.4}.
If we denote the Lorentz transformation operator to be 
$U(\Lambda)$, the equation
\begin{equation}
U(\Lambda){\cal L}({\hat x},\;{\hat\theta}^{\mu\nu})U^{-1}(\Lambda)=
{\cal L}({\hat x}{\,'},\;{\hat\theta}^{\,'}\rule{0mm}{2.8mm}^{\mu\nu})
\end{equation}
holds. Since the Lagrangian is invariant under Lorentz transformation, 
the nontrivial equation
\begin{equation}
{\cal L}({\hat x},\;{\hat\theta}^{\mu\nu})=
{\cal L}({\hat x}{\,'},\;{\hat\theta}^{\,'}\rule{0mm}{2.8mm}^{\mu\nu})
\label{3.7}
\end{equation}
follows. 
In order to obtain the Lorentz invariant Lagrangian from this equation, 
we derive several useful equations.
When the Lorentz transformation operator $U(\Lambda)$ works on Eq.\eqref{3.4}
the equation
\begin{equation}
U(\Lambda)\,{\hat\theta}^{\,\mu\nu}U^{-1}(\Lambda)\,U(\Lambda)\,|\,\theta>
=\,{\hat\theta}^{\,'}\rule{0mm}{3mm}^{\mu\nu}\,|\,\theta^{'}>
={\theta}^{\mu\nu}\,|\,\theta^{'}>
\end{equation}
 follows, where
\begin{equation}
\begin{aligned}
|\,\theta^{'}>&=U(\Lambda)\,|\,\theta>,\hskip1cm
<\theta^{'}\,|=<\theta\,|\,U^{-1}(\Lambda),\\
{\hat\theta}^{\,'}\rule{0mm}{3mm}^{\mu\nu}&=U(\Lambda)\,{\hat\theta}^{\,\mu\nu}\,U^{-1}(\Lambda).
\end{aligned}
\end{equation}
Since the Lorentz transformation for operator ${\hat\theta}^{\,\mu\nu}$ is
\begin{equation}
U(\Lambda) {\hat\theta}^{\,\mu\nu}U^{-1}(\Lambda)=
\Lambda^{\;\,\mu}_{\rho}\Lambda^{\;\,\nu}_{\sigma}
{\hat\theta}^{\,\rho\sigma}=\,{\hat\theta}^{\,'}\rule{0mm}{3mm}^{\mu\nu}
\label{3.10}
\end{equation}
the Lorentz transformation for its eigenvalue ${\theta}^{\,\mu\nu}$ is
\begin{equation}
\Lambda^{\;\,\mu}_{\rho}\Lambda^{\;\,\nu}_{\sigma}
{\theta}^{\,\rho\sigma}=\,{\theta}^{\,'}\rule{0mm}{3mm}^{\mu\nu}
\end{equation}
where ${\theta}^{\,'}\rule{0mm}{3mm}^{\mu\nu}$ is defined as 
\begin{equation}
 {\hat\theta}^{\,'}\rule{0mm}{3mm}^{\,\mu\nu}\,|\,\theta>
=\,{\theta}^{\,'}\rule{0mm}{3mm}^{\mu\nu}\,|\,\theta>.
\label{3.14}
\end{equation}

Sandwiching (3.7) between $<\theta\,|$ and $|\,\theta>$, we obtain the equation 
\begin{align}
<\theta\,|\,{\cal L}({\hat x},\,{\hat\theta})\,|\,\theta>
=<\theta\,|\,{\cal L}({\hat x}{\,'},\;{\hat\theta}^{\,'}\rule{0mm}{2.8mm}^{\mu\nu})\,|\,\theta>
\end{align}
which owing to {(3.12)} leads to 
\begin{align}
\,{\cal L}({\hat x},\,{\theta}\rule{0mm}{2.8mm}^{\mu\nu})\,
={\cal L}({\hat x}{\,'},\;{\theta}^{\,'}\rule{0mm}{2.8mm}^{\mu\nu}).
\label{3.6}
\end{align}
after the normalization factor
$<\theta\,|\,\theta>$ is scaled out.
Then, taking the trace over the spacetime operator, we obtain
\begin{align}
\int d^4x \,{\cal L}(x,\,{\theta}\rule{0mm}{2.8mm}^{\mu\nu})\,
=\int d^4x'{\cal L}({x}{\,'},\;{\theta}^{\,'}\rule{0mm}{2.8mm}^{\mu\nu}).
\end{align}
which shows that the integration of Lagrangian over $x^\mu$ is Lorentz invariant and the field theory constructed from it is Lorentz covariant.
It should be noted that Lorentz tensor $\theta^{\mu\nu}$ which 
characterizes the noncommutativity 
of spacetime is observable in experiments. It is neither an internal parameter
nor the field on NC space.
\par
For the sake of clearer understanding, we take the simple example.
It is apparent that
the equation
\begin{align}
e^{ip_{1\mu}{\hat x}^\mu}e^{ip_{2\nu}{\hat x}^\nu}
=e^{ip\,'_{1\mu}{\hat x}\,'\rule{0mm}{2.5mm}^\mu}e^{ip\,'_{2\nu}{\hat x}\,'\rule{0mm}{2.5mm}^\nu}\label{A315}
\end{align}
shows Lorentz invariance since both $p_\mu$ and ${\hat x}^\mu$ are Lorentz 
vectors.
According to the math formula, the left hand side of 
\eqref{A315} is changed  as follows
\begin{align}
e^{ip_{1\mu}{\hat x}^\mu}e^{ip_{2\nu}{\hat x}^\nu}
&=e^{ip_{1\mu}{\hat x}^\mu+ip_{2\nu}{\hat x}^\nu}
e^{-\frac12p_{1\mu}p_{2\nu}[{\hat x}^\mu,\,{\hat x}^\nu]}\nonumber\\
&=e^{ip_{1\mu}{\hat x}^\mu+ip_{2\nu}{\hat x}^\nu}
e^{-i\frac12p_{1\mu}p_{2\nu}{\hat\theta}^{\mu\nu}}.
\end{align}
Then, from \eqref{A315}, the equality
\begin{align}
e^{ip_{1\mu}{\hat x}^\mu+ip_{2\nu}{\hat x}^\nu}
e^{-i\frac12p_{1\mu}p_{2\nu}{\hat\theta}^{\mu\nu}}
=e^{ip\,'_{1\mu}{\hat x}\,'\rule{0mm}{2.5mm}^{\mu}+ip\,'_{2\nu}{\hat x}\,'\rule{0mm}{2.5mm}^{\nu}}
e^{-i\frac12p\,'_{1\mu}p\,'_{2\nu}{\hat\theta}{\,'}\rule{0mm}{2.5mm}^{\mu\nu}}
\label{B316}
\end{align}
follows. By sandwiching \eqref{B316} in $<\theta\,|$ and $|\,\theta>$, we obtain
\begin{align}
<\theta\,|e^{ip_{1\mu}{\hat x}^\mu+ip_{2\nu}{\hat x}^\nu}
e^{-i\frac12p_{1\mu}p_{2\nu}{\hat\theta}^{\mu\nu}}|\,\theta>
=<\theta\,|e^{ip\,'_{1\mu}{\hat x}\,'\rule{0mm}{2.5mm}^{\mu}+ip\,'_{2\nu}{\hat x}\,'\rule{0mm}{2.5mm}^{\nu}}
e^{-i\frac12p\,'_{1\mu}p\,'_{2\nu}{\hat\theta}{\,'}\rule{0mm}{2.5mm}^{\mu\nu}}
|\,\theta>.
\end{align}
Owing to equations derived from (3.4) and ({3.12})
\begin{align}
&<\theta\,|\hat\theta^{\mu\nu}|\,\theta>
=\theta^{\mu\nu}<\theta\,|\,\theta>,\hskip1cm
<\theta\,|{\hat\theta}{\,'}\rule{0mm}{2.5mm}^{\mu\nu}|\,\theta>
={\theta}{\,'}\rule{0mm}{2.5mm}^{\mu\nu}<\theta\,|\,\theta>,
\end{align}
the equation
\begin{align}
e^{ip_{1\mu}{\hat x}^\mu+ip_{2\nu}{\hat x}^\nu}
e^{-i\frac12p_{1\mu}p_{2\nu}{\theta}^{\mu\nu}}=
e^{ip\,'_{1\mu}{\hat x}\,'\rule{0mm}{2.5mm}^{\mu}+ip\,'_{2\nu}{\hat x}\,'\rule{0mm}{2.5mm}^{\nu}}
e^{-i\frac12p\,'_{1\mu}p\,'_{2\nu}{\theta}{\,'}\rule{0mm}{2.5mm}^{\mu\nu}}
\label{319}
\end{align}
follows because the factor $<\theta\,|\,\theta>$ in both sides is scaled out.
 Since the trace over the NC spacetime is given by
\begin{align}
\text{Tr}e^{ip_\mu {\hat x}^{\mu}}=(2\pi)^4\delta^4(p)=\int e^{ip_\mu {x}^\mu}
d^4x,
\end{align}
the trace of \eqref{319} with respect to the space-time operator
is
\begin{align}
\int d^4x\; e^{ip_{1\mu}{ x}^\mu+ip_{2\nu}{x}^\nu}
e^{-i\frac12p_{1\mu}p_{2\nu}{\theta}^{\mu\nu}}=\int d^4x'\;
e^{ip\,'_{1\mu}{ x}\,'\rule{0mm}{2.5mm}^{\mu}+ip\,'_{2\nu}{ x}\,'\rule{0mm}{2.5mm}^{\nu}}
e^{-i\frac12p\,'_{1\mu}p\,'_{2\nu}{\theta}{\,'}\rule{0mm}{2.5mm}^{\mu\nu}}.
\end{align}
which leads to the equation
\begin{align}
\int d^4x\; e^{ip_{1\mu}{ x}^\mu}\ast e^{ip_{2\nu}{x}^\nu}
=\int d^4x'\;
e^{ip\,'_{1\mu}{ x}\,'\rule{0mm}{2.5mm}^{\mu}}\ast e^{ip\,'_{2\nu}{ x}\,'\rule{0mm}{2.5mm}^{\nu}}.
\end{align}
Similarly, we can obtain
\begin{align}
\int d^4x\; e^{ip_{1\mu}{ x}^\mu}\ast e^{ip_{2\nu}{x}^\nu}\ast e^{ip_{3\sigma}{x}^\sigma}
=\int d^4x'\;
e^{ip\,'_{1\mu}{ x}\,'\rule{0mm}{2.5mm}^{\mu}}\ast e^{ip\,'_{2\nu}{ x}\,'\rule{0mm}{2.5mm}^{\nu}}\ast e^{ip\,'_{3\sigma}{ x}\,'\rule{0mm}{2.5mm}^{\sigma}}.
\end{align}

\par
Since Lagrangian is written in terms of field operators, 
we explain the more realistic case.
When
$f({\hat x})\,g({\hat x})$
is Lorentz invariant, the equation
\begin{align}
f({\hat x})\,g({\hat x})=f\,'({\hat x}\,')\,g\,'({\hat x}\,')
\label{A324}
\end{align}
follows. For example,
$f({\hat x})\,g({\hat x})$ may take the form
\begin{align}
f({\hat x})\,g({\hat x})=F^{\mu\nu}({\hat x})F_{\mu\nu}({\hat x}).
\end{align}
In the equation \eqref{A324}, $f\,'({\hat x}\,')$ 
and $g\,'({\hat x}\,')$
are Lorentz transformations of $f({\hat x})$ and $g({\hat x})$, respectively. 
\begin{align}
&f\,'({\hat x}\,')=U(\Lambda)f({\hat x})U^{-1}(\Lambda),\\
&g\,'({\hat x}\,')=U(\Lambda)g({\hat x})U^{-1}(\Lambda).
\end{align}
Then, sandwiching \eqref{A324} between $<\theta\,|$ and $|\,\theta>$ and taking the trace over 
spcetime operator, the left hand side of \eqref{A324} is changed to
\begin{align}
\text{Tr}<\theta\,|f({\hat x})\,g({\hat x})\,|\,\theta>&=\int d^4p_1\int\,d^4p
\,F(p_1)\,G(p_2)
\text{Tr}<\theta\,|e^{ip_{1\mu}{\hat x}^\mu}e^{ip_{2\nu}{\hat x}^\nu}
\,|\,\theta>\nonumber\\
&=\int d^4x\,e^{i\frac12\partial_{1\mu}\partial_{2\nu}{\theta}^{\mu\nu}}f(x_1)\,g(x_2)
\left.\rule{0mm}{5mm}\right|_{x_1=x_2=x},
\end{align}
which together with \eqref{A324} leads to
\begin{align}
\int d^4x\,f(x)\ast g(x)=
\int d^4x'\,f\,'(x')\ast g\,'(x')
\label{A3.29}
\end{align}
which concludes that 
the integration over $x^\mu$ is Lorentz invariant  even if it is not integrated over $\theta$.
It should be noted that even if $f(x)=f_1(x)\ast f_2(x)$, \eqref{A3.29}
is true.

\par
In this stage, the choice of $\theta^{\,\mu\nu}$ is arbitrary. 
However, quantization restricts the allowable region of $\theta^{\,\mu\nu}$
because when the timelike noncommutativity $\theta^{0i}\ne0$ exists,
the conjugate momentum of a filed $\phi$ defined by
\begin{equation}
{\Pi}=\frac{\partial {\cal L}}{\partial(\partial^0 \phi)}
\end{equation}
is not qualified as an appropriate momentum owing to the infinite
series of time derivatives in the Moyal $\ast$products.
Thus, we can restrict the region of $\theta^{\,\mu\nu}$ in such a way that
we can render 
$\theta^{\,0i}$ to be 0 by making an appropriate Lorentz transformation of 
$\theta^{\,\mu\nu}$.
The unitarity problem pointed out by Gomis and Mehen \cite{GM}
may vanish by considering the Lorentz invariance of the theory 
and the proper choice of $\theta^{\,\mu\nu}$ as discussed above. 

\section{Conclusion}
Lorentz covariance is the fundamental principle of every relativistic field theory which insures consistent physical descriptions. Even if the space-time is noncommutative, field theories on it should keep Lorentz covariance.
In this paper, we construct the Lorentz invariant action on NC spacetime 
\begin{align}
S=\text{Tr}<\theta\,|\,{\cal L}({\hat x},\,{\hat\theta})\,|\,\theta>
=\int d^4x \,{\cal L}(x,\,{\theta}\rule{0mm}{2.8mm}^{\mu\nu})
\end{align}
which leads to the Lorentz covariant field theory on NC spacetime.
\par
We can choose the spacelike noncommutativity ($\theta^{0i}=0$) after 
the appropriate Lorentz transformation and so any serious outcomes such as 
quantization and unitarity violation don't appear.
As stated before, it should be stressed that Lorentz tensor $\theta^{\mu\nu}$ which characterizes the noncommutativity 
of spacetime is observable in experiments.

\par



\begin{thebibliography}{99}
\bibitem{SJ}
M. M. Sheikh-Jabbari, Phys. Lett. {\bf 425}, 48 (1998);\\
F. Ardalan, H. Arfaei, M. M. Sheikh-Jabbari, JHEP, {\bf 9902}, 016 (1999),
Nucl. Phys. {\bf 576}, 578 (2000).\\
C-S. Chi, P-H. Ho, Nucl. Phys. {\bf 550}, 578 (1999);
Nucl. Phys. {\bf 568}, 477 (2000).
\bibitem{DH}
M. R. Douglas, C. Hull, JHEP, {\bf 9802}, 008 (1998).\\
M. M. Sheikh-Jabbari, Phys. Lett. {\bf 450}, 119 (1999)
\bibitem{SW}
N. Seiberg and E. Witten, JHEP {\bf 9909}, 032 (1999).
\bibitem{Kad} V. G. Kadysshevskii, 
Sov. Phys. JETP {\bf 14}, 1340 (1962).
\bibitem{Filk} T. FIlk,
Phys. Letters, {\bf 376}, 53 (1996).\\
S. Minwa, M. Van Raamsdonk, N. Seiberg, JHEP {\bf 0002}, 020 (2000).\\
I. Ya. Aref'eva, D. M. Belov, A. S. Koshelev, Phys. Letters, {\bf B476},
431 (2000). 
\bibitem{SST} N. Seiberg, L. Susskind, N. Toumbas, JHEP, {\bf 0006},
044 (2000).
\bibitem{Micu}
A. Micu, M. M. Sheikh-Jabbari,  JHEP {\bf 0101}, 025 (2001). 
\bibitem{DFR}
S. Doplicher, K. Fredenhagen and J. E. Roberts, Phys. Lett. {\bf B331}, 39 (1994) ; Commun. Math. Phys. {\bf 172}, 187 (1995).
\bibitem{CCZ}
C. E. Carlson, C. D. Carone and N. Zobin, Phys. Rev. {\bf D 66}, 075001 (2002).
\bibitem{Hayakawa} M. Hayakawa, Phys. Lett. {\bf B478}, 394 (2000).
\bibitem{Armoni} A. Armoni, Nucl. Phys. {\bf B593}, 229 (2001).
\bibitem{KMOU}
H. Kase, K. Morita, Y. Okumura and E. Umezawa, Prog. Theor. Phys. {\bf 109},
663 (2003).
\bibitem{GM}
J. Gomis and T. Mehen, Nucl. Phys. {\bf B591}, 178 (2000).
\end{thebibliography}
\end{document}